\documentclass[aps,pra,superscriptaddress,amsmath,amssymb,preprintnumbers,showpacs,twocolumn]{revtex4}
\usepackage{amssymb}
\usepackage{graphicx}
\usepackage{bm}
\usepackage{subfigure}
\usepackage{color}

\newcommand{\Ignore}[1]{}

\newcommand{\Ket}[1]{\left\vert #1\right\rangle}
\newcommand{\Bra}[1]{\left\langle #1\right\vert}

\newcommand{\BraKet}[2]{\left\langle#1\vert #2\right\rangle}
\newcommand{\KetBra}[2]{\left\vert#1\right\rangle\left\langle#2\right\vert}

\renewcommand{\eqref}[1]{Eq.~(\ref{#1})}

\def\e{\mathrm{e}}
\def\ii{\mathrm{i}}
\def\LZSM{LZSM}

\newcommand{\LZ}{Landau-Zener-Stueckelberg-Majorana }

\begin{document}

\title{Dynamics of a two-state system through a real level crossing}

\author{Benedetto D. Militello}
\address{Dipartimento di Fisica e Chimica, Universit\`a degli Studi di Palermo, Via Archirafi 36, I-90123 Palermo, Italy}
\author{Nikolay V. Vitanov}
\affiliation{Department of Physics, St Kliment Ohridski University
of Sofia, 5 James Bourchier Blvd, 1164 Sofia, Bulgaria}

\begin{abstract}
The dynamics of a two-state system whose energies undergo a real
crossing at some instant of time is studied. At this instant, both
the coupling and the detuning vanish simultaneously, which leads
to an exact degeneracy of the eigenenergies of the system. It is
found that the dynamics of the system is primarily determined by
the manner in which the degeneracy occurs. This interesting
behavior is reminiscent of a symmetry breaking process, since the
totally symmetric situation occurring at the crossing is
significantly altered by infinitesimal quantities, which remove
the degeneracy, with very important dynamical implications from
there on. A very simple analytical formula is derived, which is
found to describe the population changes very accurately.
\end{abstract}


\maketitle

\section{Introduction}

In the realm of adiabatic evolution~\cite{ref:Messiah}, the
dynamics of a two-state system whose energies cross at some point
plays an important role. The exactly solvable model of \LZ
(\LZSM)~\cite{ref:LZSM-1,ref:LZSM-2,ref:LZSM-3,ref:LZSM-4} is the
most famous example of such class of problems. Its widespread
applications range from quantum optics and atomic physics to
chemistry and biophysics. In fact, this model is of interest in
several physical systems~\cite{ref:Nori_review}, such as Josephson
junctions~\cite{ref:JJ1,ref:JJ2}, cold atoms in optical
lattices~\cite{ref:BEC} and spinorial Bose-Einstein condensates
\cite{ref:Nori_spinorial}. In order to better consider realistic
situations, the implications of having a finite
duration~\cite{ref:Vitanov97b,ref:Vitanov99} or a nonlinear
crossing~\cite{ref:Vitanov99nonlinear} in the experiments have
also been studied. Generalizations of the original model to the
cases wherein more than two states are involved have been
considered \cite{ref:Multilevel1,ref:Multilevel2,ref:Multilevel3},
as well as level crossings occurring in systems described by
nonlinear equations~\cite{ref:Nonlinear}. The relevance of the
\LZSM~model in statistical physics, especially in connection with
phase transitions in the Ising model, has been
discussed~\cite{ref:Chains1,ref:Chains2}. Moreover, recently, a
geometric Landau-Zener-St\"uckelberg interferometer has been
realized in single trapped ions by Zhang {\it et al}
~\cite{ref:Zhang2014} on the basis of a proposal by Gasparinetti
{\it et al}~\cite{ref:Gasparinetti2011}, and Landau-Zener
transitions in frozen pairs of Rydberg atoms have been observed by
Saquet {\it et al}~\cite{ref:Saquet2010}.

In the original \LZSM~model, the bare energy levels of a two-state
system cross at some instant of time, but, due to the presence of
a (constant) interaction, the degeneracy is removed in the
adiabatic eigenenergies. Then the adiabatic following of an
instantaneous eigenstate of the Hamiltonian leads to a perfect
transition from one bare state to the other. Beyond the perfect
adiabaticity, the correction to the dynamics is given by the
formula found by Landau, Zener, St\"uckelberg and Majorana which
expresses the transition probability between the eigenstates
(non-adiabatic transitions) of the
Hamiltonian~\cite{ref:LZSM-1,ref:LZSM-2,ref:LZSM-3,ref:LZSM-4},
and which is equal to the no-transition probability between the bare
states.

In 1990 Fishman {\it et al} have introduced a variant of the
original \LZSM~model where neither the bare energies nor the
dressed ones really cross~\cite{ref:Fishman1990}, showing that,
even in the presence of the same dressed energies, the
circumstance that the bare energies do not cross as in the
standard model determines significatively different results. One
of the most interesting experimental realizations of this modified
\LZSM~model has been provided by Bouwmeester {\it et
al}~\cite{ref:Bouwmeester1995}, who have observed such modified
Landau-Zener dynamics (they call the relevant mathematical
description the \lq hidden-crossing model\rq) in classical optical
systems. By the way, it is worth mentioning that with their
experiment Bouwmeester {\it et al} have demonstrated that the
validity of the \LZSM\, model (and related ones) is not limited to
the realm of quantum mechanics.

The variant of the \LZSM~model complementary to the
hidden-crossing model corresponds to a physical situation where
both bare and dressed energies cross at the same time, which has
never been studied. It is then of interest to analyze the
adiabatic evolution of a two-state system in the special case
where a real degeneracy occurs at the crossing of the bare
energies. In other words, we want to study the case where at the
same time when the bare energies cross, the interaction, assumed
non vanishing in the rest of the process, vanishes too. At this
instant the adiabatic (i.e., dressed) energies, just like the
diabatic (i.e., bare) energies, become degenerate. In this case
the \LZSM~formula is not applicable, and in order to estimate the
transition probability we have to scrutinize the intricate
behavior of the eigenenergies and the nonadiabatic coupling in the
vicinity of this real crossing.

The paper is organized as follows. In the next section, we
introduce the model Hamiltonian we are going to analyze. In
section~\ref{sec:adiabaticity} we discuss the condition for the
validity of the adiabatic approximation, firstly in general and
secondly focusing on the instant of time when the level crossing
occurs. In section~\ref{sec:diabatic_transitions} we report on an
analysis that allows to predict the efficiency of population
transfer on the basis of the peculiarities of the Hamiltonian.
Finally, in section~\ref{sec:discussion} we summarize the results
and provide some conclusive remarks.

\section{The Model}\label{sec:TheModel}

Consider a two-state system whose dynamics is governed by a Hamiltonian having the following form ($\hbar=1$):
\begin{equation}
H = \frac{\Delta}{2}\sigma_z + \frac{\Omega}{2} \sigma_x = %
\frac{1}{2}\,\left(
\begin{array}{cc}
-\Delta & \Omega \cr %
\Omega & \Delta \cr %
\end{array}
\right)\,.
\end{equation}
This Hamiltonian has eigenvalues
\begin{equation}
  E_\pm \equiv \omega_\pm = \pm\frac{\Omega}{2}\sqrt{1+\alpha^2}\,,
\end{equation}
and eigenstates
\begin{subequations}
\begin{align}
  \Ket{\psi_+} &= \cos\theta \Ket{\downarrow} + \sin\theta \Ket{\uparrow}\,,\\
  \Ket{\psi_-} &= \sin\theta \Ket{\downarrow} - \cos\theta \Ket{\uparrow}\,,
\end{align}
\end{subequations}
where
\begin{subequations}
\begin{align}  \label{eq:ThetaDef}
  \tan\theta &= \alpha + \sqrt{1+\alpha^2}\,,\\
  \alpha &= \frac{\Delta}{\Omega}\,.
\end{align}
\end{subequations}

In the \LZSM~model, the interaction is constant ($\dot\Omega=0$)
and the bare energies change linearly ($\Delta=\kappa t$). If the
change is very slow ($\kappa$ small enough) then the evolution is
adiabatic. Moreover, when $t$ varies from $-\infty$ to $+\infty$,
the dressed states change from one bare state to the other; for
example, $\Ket{\psi_+(-\infty)}=\Ket{\downarrow}$ becomes
$\Ket{\psi_+(\infty)}=\Ket{\uparrow}$). Therefore, if the system
is prepared in one of the bare states, and the system follows the
corresponding dressed state, then in the end of the process each
bare state is perfectly mapped to the other bare state: population
inversion has occurred. Perfect adiabatic evolution does not exist
(except for constant Hamiltonian or for $\Delta\equiv 0$), and the
\LZSM~formula gives the deviation from perfect population
inversion.

In the following, we will instead study the adiabatic evolution
when $\Omega$ and $\Delta$ are such that $\Delta\rightarrow 0$ and
$\Omega\rightarrow 0$, when $t\rightarrow 0$, so that a real level
crossing occurs at the time instant $t=0$. Of course, there are
several ways the two quantities can approach zero, and depending
on this, the evolution of the system can be very different.

In order to maintain the possibility of having a complete
population transfer, we will always assume that $\alpha$ varies
from $-\infty$ (i.e., a huge negative value) to $+\infty$ (a huge
positive value) or vice versa. Indeed, this will make each of the
dressed states change from being exactly (i.e., essentially) equal
to one bare state to being exactly (i.e., essentially) equal to
the other bare state along the experiment.

\section{Adiabatic Evolution}\label{sec:adiabaticity}

After introducing the unitary operator describing the passage to
the basis of the instantaneous eigenstates of the Hamiltonian,
\begin{eqnarray}
\nonumber T_\mathrm{A} &=&
\KetBra{\psi_+(-\infty)}{\psi_+(t)}+\KetBra{\psi_-(-\infty)}{\psi_-(t)} \\
&=& \KetBra{\downarrow}{\psi_+(t)}+\KetBra{\uparrow}{\psi_-(t)}
\,,
\end{eqnarray}
we easily find that in this basis the evolution is generated by
the following operator:
\begin{equation}
-\ii T_\mathrm{A} \dot{T}_\mathrm{A}^\dag + T_\mathrm{A} H
T_\mathrm{A}^\dag = \left(
\begin{array}{cc}
\omega_+ & \ii\dot\theta \cr %
-\ii\dot\theta & \omega_- \cr %
\end{array}
\right)\,.
\end{equation}
It is worth noting that for $\dot\theta = 0$ one would seemingly
have perfect adiabatic evolution (in fact, each of the two
instantaneous eigenstates would just acquire a phase, without
undergoing any transitions). But, if $\dot\theta = 0$ then there
is no change of the eigenstates of the Hamiltonian, and then there
is no real adiabatic evolution. In fact, the adiabatic following
of the eigenstates of the Hamiltonian governing the system is
intrinsically the result of some approximations and obtaining an
exact adiabatic following of the eigenstates of the system
Hamiltonian is a \textit{chimera}. There is, seemingly, a
violation to such an assertion given by the shortcuts to
adiabaticity~\cite{ref:shortcuts1,ref:shortcuts2,ref:shortcuts3}.
In fact, such techniques
 allow to realize the exact adiabatic
following of the eigenstates of a given Hamiltonian. But the
crucial point of such shortcuts is that the exact following is
obtained through an additional interaction, so that, in the end,
the adiabatic evolution does not coincide with the adiabatic
following of the eigenstates of the system Hamiltonian. In other
words, in order to realize the adiabatic following of the
eigenstates of a given Hamiltonian $H(t)$ it is necessary to let
the system evolve under the action of $H(t)+H_{sc}(t)$, where
$H_{sc}(t)$ is a suitable additional term that allows to realize
the shortcut and which does not commute with $H(t)$. In this way,
there is no following of the eigenstates of the total Hamiltonian
of the system $H(t)+H_{sc}(t)$.

\subsection{General condition for adiabaticity}

In order to have an adiabatic evolution the terms responsible for
transitions between the eigenstates of the Hamiltonian must be
negligible, which in our case means: $\eta \equiv |\dot\theta| /
|\omega_+-\omega_-| \ll 1$. Now, since
\begin{equation}
\dot\theta =
\frac{\alpha+\sqrt{1+\alpha^2}}{1+(\alpha+\sqrt{1+\alpha^2})^2}
\,\frac{\dot\alpha}{\sqrt{1+\alpha^2}} \,,
\end{equation}
one finds:
\begin{eqnarray}
\nonumber
\eta &=& |\dot\theta|\,/\, |\omega_+-\omega_-| \\
&=& \label{eq:eta_condition_alpha}
\left|\frac{\alpha+\sqrt{1+\alpha^2}}{1+(\alpha+\sqrt{1+\alpha^2})^2}\right|
\,\times\left|\frac{\dot\alpha}{\Omega(1+\alpha^2)}\right| \,,
\end{eqnarray}
which can also be cast in the following form (consider that \,
$\alpha+\sqrt{1+\alpha^2}=1/(-\alpha+\sqrt{1+\alpha^2})$ and then
$2\sqrt{1+\alpha^2}=\tan\theta + 1/\tan\theta$):
\begin{equation}
\eta=\left| \, \sin(2\theta) \right|^3 \times \left|\,\dot\alpha /
\Omega \right| \,.
\end{equation}

Putting $\eta$ in this form, it is clear that a sufficient
condition to guarantee an adiabatic evolution is
$\left|\,\dot\alpha / \Omega \right| \ll 1$. Nevertheless, in some
cases, a finite or even a divergent value of $\left|\,\dot\alpha /
\Omega \right|$ can be compensated by the smallness of
$\sin(2\theta)$.

Let us come back to \eqref{eq:eta_condition_alpha} and consider
what the condition for adiabaticity becomes in the cases
$|\alpha|\ll 1$, $\alpha\sim 1$ and $|\alpha|\gg 1$. First of all,
consider that both for $|\alpha|\ll 1$ and $\alpha\sim 1$ one can
say that $\eta \sim \left|\dot\alpha / \Omega\right| =
|(\dot\Omega\Delta-\Omega\dot\Delta)/\Omega^3|$, while for
$|\alpha|\gg 1$ one has $\eta \sim \left|\dot\alpha /
(\alpha^3\Omega)\right| =
|(\dot\Omega\Delta-\Omega\dot\Delta)/\Delta^3|$. Therefore, the
conditions to have an adiabatic evolution become:
\begin{subequations}
\begin{eqnarray}
\left\{
\begin{array}{l}
|\alpha|\ll 1 \\
|\alpha|\sim 1
\end{array}
\right. &\Rightarrow &
|(\dot\Omega\Delta-\Omega\dot\Delta)/\Omega^3| \ll 1\,, \\
|\alpha|\gg 1 \,\, &\Rightarrow & \,\,
|(\dot\Omega\Delta-\Omega\dot\Delta)/\Delta^3| \ll 1\,.
\end{eqnarray}
\end{subequations}

Note that when $|\alpha| \gg 1$, for $\alpha>0$ one has
$\theta\rightarrow \pi/2$, while for $\alpha<0$ it turns out
$\theta\rightarrow 0$, and then $\sin(2\theta)\rightarrow 0$ in
both cases. This is just the case mentioned above where smallness
of $\sin(2\theta)$ can compensate higher values of
$|\dot\alpha/\Omega|$, and this is the reason why the condition
for having an adiabatic evolution assumes a different form.

\subsection{Adiabaticity at the level crossing}

It is of interest to give some general indication of the validity
of the adiabatic approximation close to the level crossing. In
order to develop such an analysis, let us consider the behavior of
$\Delta(t)$ and $\Omega(t)$ close to the level crossing occurring
at $t=0$, assuming that the leading terms in their Taylor
expansions versus $t$ at the crossing ($t=0$) are
\begin{subequations}
\begin{eqnarray}
\Delta \simeq A t^a \,, \\
\Omega \simeq B t^b \,,
\end{eqnarray}
\end{subequations}
with $a,b > 0$. Then, in the case $\alpha\ll 1$ ($a>b$), one has:
\begin{subequations}
\begin{equation}
\eta \sim \left|\frac{A(b-a)}{B^2}\right| \times |t|^{a-2b-1} \,, \\
\end{equation}
and to guarantee $\eta\ll 1$ one needs $a>2b+1$. 

In the case $\alpha\gg 1$ ($b>a$) it turns out,
\begin{equation}
\eta \sim \left|\frac{A(b-a)}{B^2}\right| \times |t|^{b-2a-1} \,.
\end{equation}
\end{subequations}
and to guarantee $\eta\ll 1$ one needs $a<(b-1)/2$. 

Therefore, each of the following inequalities guarantee adiabaticity in
the very neighborhood of $t=0$:
\begin{subequations}
\begin{align}
a &> 2b + 1  \,,\\
b &> 2a + 1  \,.
\end{align}
\end{subequations}

The last case to consider is given by the situation $|\alpha|\sim
1$, which deserves a specific analysis since it requires the
analysis of further terms in the Taylor expansion of both $\Delta(t)$
and $\Omega(t)$.

\section{Nonadiabatic transitions}\label{sec:diabatic_transitions}

In this section we give some general dynamical properties that
allow to forecast the efficiency of the population transfer on the
basis of an analysis of $\Delta(t)$ and $\Omega(t)$ at the level
crossing.

\subsection{Transitions at the level crossing}

Let us first of all assume that the evolution of the system is
essentially adiabatic everywhere, with the only possible exception
of the very neighborhood of $t=0$. The total evolution operator
(that is, the evolution operator in the Schr\"odinger picture) can
be written as:
\begin{eqnarray}
\nonumber
U(+\infty,-\infty) &=& U(+\infty,0^+) U(0^+,0^-)  U(0^-,-\infty)\\
\nonumber
&\approx& U_\mathrm{A}(+\infty,0^+) U(0^+,0^-)
U_\mathrm{A}(0^-,-\infty)\,, \\
\end{eqnarray}
where we have introduced the adiabatic evolution operator:
\begin{eqnarray}
U_\mathrm{A}(t_2,t_1) = \sum_{k=\pm} e^{-\ii
\varphi_k(t_2,t_1)}\KetBra{\psi_k(t_2)}{\psi_k(t_1)}\,,
\end{eqnarray}
with
\begin{eqnarray}
\varphi_k(t_2,t_1) = \int_{t_1}^{t_2} \left( \omega_k(s) - \ii
\BraKet{\psi_k(s)}{\dot{\psi}_k(s)} \right) \mathrm{d}s\,.
\end{eqnarray}

Now, since around $t=0$ the Hamiltonian $H$ does not exhibit
singularities (far from this, it is infinitesimal), one has
$U(0^+,0^-)\approx \mathbf{I}$, and then:
\begin{eqnarray}
U(+\infty,-\infty) &\approx& U_\mathrm{A}(+\infty,0^+)
U_\mathrm{A}(0^-,-\infty)\,,%
\end{eqnarray}
from which:
\begin{eqnarray}
U(+\infty,-\infty) &=& \sum_{k=\pm} \sum_{l=\pm} T_{kl}
\Ket{\psi_k(+\infty)}\Bra{\psi_l(-\infty)} \,,
\end{eqnarray}
where
\begin{subequations}
\begin{align}
T_{kl} &= \e^{- \ii (\varphi_k^-+\varphi_l^+)} \BraKet{\psi_k(0^+)}{\psi_l(0^-)} \,,\\
\varphi_k^- &= \varphi_k(0^-,-\infty)\,, \\ 
\varphi_l^+ &= \varphi_k(+\infty,0^+)\,.    
\end{align}
\end{subequations}

Now, consider what follows:
\begin{subequations}
\begin{align}
\BraKet{\psi_\pm(s)}{\dot{\psi}_\pm(s)} &= 0 \,, \\
\BraKet{\psi_\pm(s)}{\dot{\psi}_\mp(s)} &= \pm \dot\theta \,,
\end{align}
\end{subequations}
and, more important,
\begin{subequations}\label{eq:MatrEqs}
\begin{align}
\label{eq:MatrEl1}
\BraKet{\psi_+(0^+)}{\psi_+(0^-)} &= \cos\Theta\,, \\
\label{eq:MatrEl2}
\BraKet{\psi_-(0^+)}{\psi_-(0^-)} &= \cos\Theta\,, \\
\label{eq:MatrEl3}
\BraKet{\psi_+(0^+)}{\psi_-(0^-)} &= -\sin\Theta\,, \\
\label{eq:MatrEl4} %
\BraKet{\psi_-(0^+)}{\psi_+(0^-)} &=\sin\Theta\,,
\end{align}
\end{subequations}
with
\begin{equation}
\Theta = \theta(0^+)-\theta(0^-)\,.
\end{equation}
From Eqs.~(\ref{eq:MatrEqs}) we can derive the probabilities of the nonadiabatic transitions.
The failure of adiabatic passage in the overall process can be quantified through the parameter $\Theta$:
\begin{subequations}
\begin{align}
\label{eq:survprob}
\left| \Bra{\psi_\pm(+\infty)} U(+\infty,-\infty) \Ket{\psi_\pm(-\infty)} \right|^2 &= \cos^2\Theta\,,\\
\label{eq:transprob} \left| \Bra{\psi_\pm(+\infty)}
U(+\infty,-\infty) \Ket{\psi_\mp(-\infty)} \right|^2 &=
\sin^2\Theta\,.
\end{align}
\end{subequations}

The first of such two equations, \eqref{eq:survprob}, says that
each of the two eigenstates of $H$, $\Ket{\psi_+(t)}$ or
$\Ket{\psi_-(t)}$, follows the adiabatic change of the Hamiltonian
with probability $\cos^2\Theta$. On the other hand,
\eqref{eq:transprob} says that each of such two states undergoes a
transition to the other state with probability $\sin^2\Theta$.

The nonadiabatic transitions are determined by the abrupt change
of the eigenstates of the Hamiltonian at $t=0$. It is interesting
to point out that, from the dynamical point of view, nothing really
happens around $t=0$, because the Hamiltonian is essentially zero
in that time interval. Nevertheless, what happens from there on is
mainly determined by the possible abrupt change of the Hamiltonian
eigenstates occurring at $t=0$.

\begin{figure}[tb]
\subfigure[]{\includegraphics[width=0.35\textwidth, angle=0]{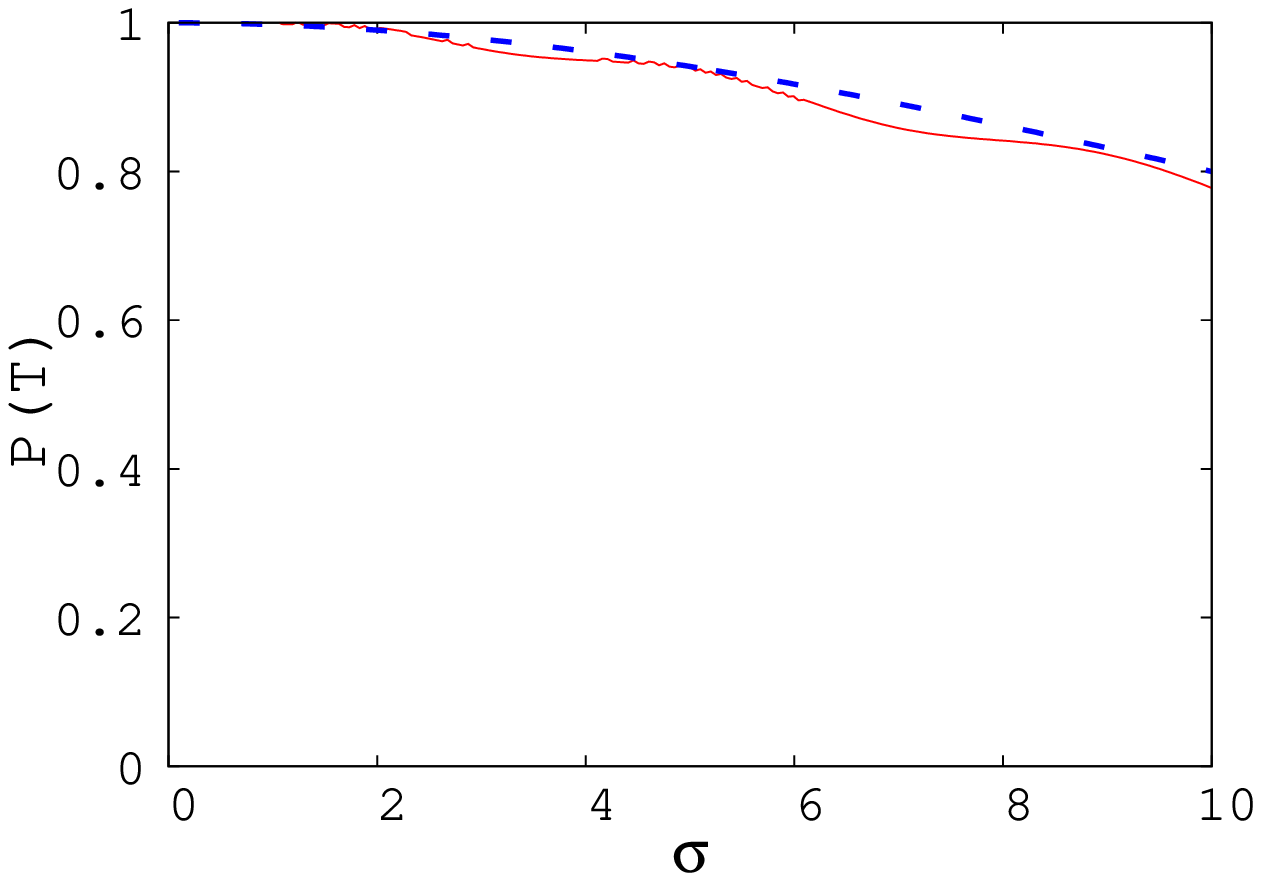}} %
\subfigure[]{\includegraphics[width=0.35\textwidth, angle=0]{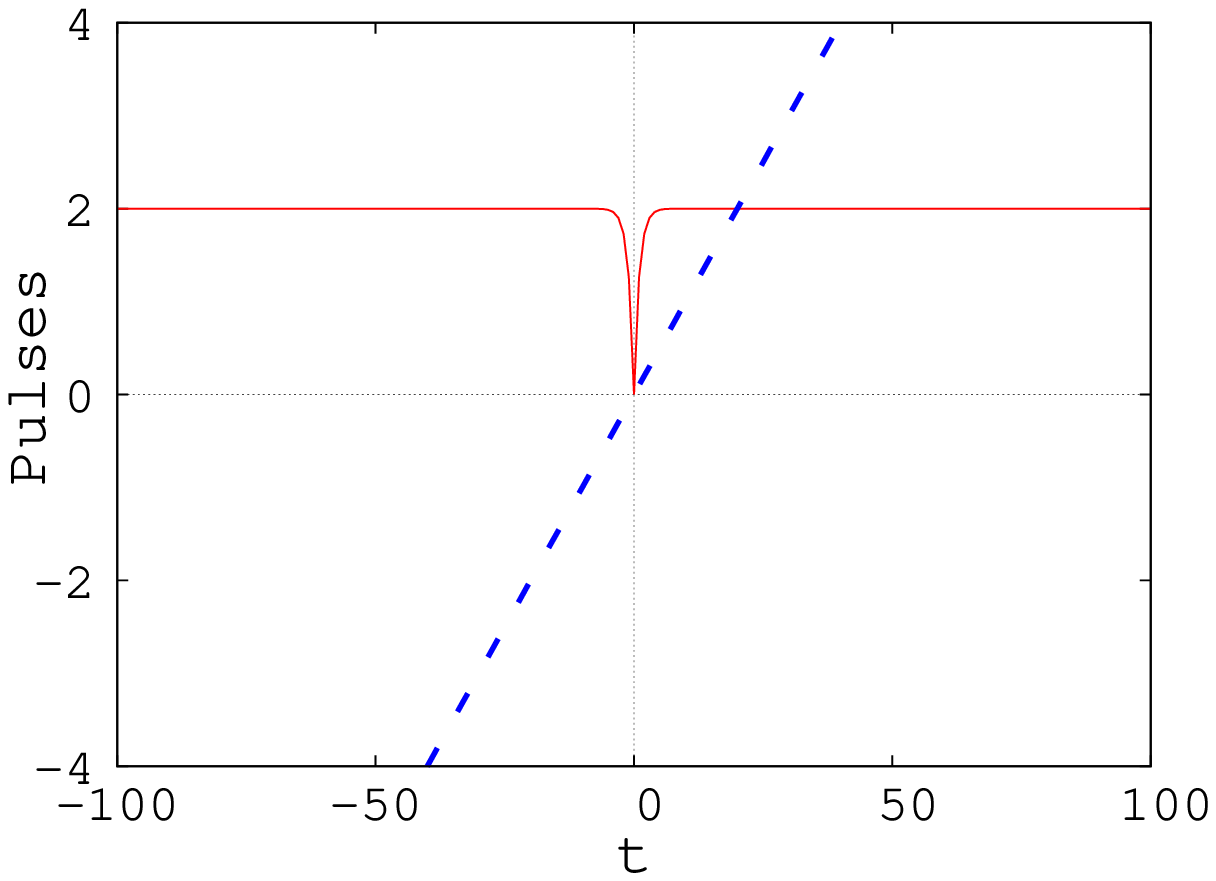}} %
\caption{(Color online) Frame (a) shows the asymptotic populations
of the bare state $\Ket{\uparrow}$ (red solid line) and the
theoretical prediction of the efficiency (blue dashed line),
$\cos^2\Theta$, vs the parameter $\sigma$ (in units of
$2/\Omega_0$). Here
$\Omega=\Omega_0(1-\mathrm{exp}(-|t/\sigma|))$, $\Delta=\kappa t$,
with $k/\Omega_0^2 = 0.025$ and $\kappa T/\Omega_0=50$, where $2T$ is the
duration of the experiment. Frame (b) shows the coupling $\Omega$
(red solid line) and the detuning $\Delta$ (blue dashed line) for
$\sigma=2/\Omega_0$. The amplitudes of the pulses are in units of
$\Omega_0/2$ and time is in units of $2/\Omega_0$.}
\label{fig:Omega_Exp}
\end{figure}

The case $\Theta=0$ can be obtained when $\alpha \rightarrow 0$,
which implies $\theta(0^+)=\theta(0^-)=\pi/4$. On the contrary,
the case $\Theta=\pi/2$ ($-\pi/2$) can be obtained when
$\alpha\rightarrow -\infty$ ($+\infty$) on the left of $t=0$ and
$\alpha\rightarrow +\infty$ ($-\infty$) on the right. Therefore,
we can say that when $\Delta$ is infinitesimal of higher order
than $\Omega$ there is no abrupt change of the eigenstates and
then there is a total transfer of population. On the other hand,
when $\Omega$ is infinitesimal of higher order than $\Delta$ we
expect an abrupt change of the eigenstates and then a very low
population transfer efficiency. When $\Delta$ and $\Omega$ are
infinitesimals of the same order, a specific analysis of the
behavior of $\theta$ is necessary.

All these results have been derived under the assumption that the
adiabatic approximation is valid everywhere far from $t=0$, and
that a possible critical behavior can occur only when the levels
cross ($t=0$).

\begin{figure}[tb]
\subfigure[]{\includegraphics[width=0.35\textwidth, angle=0]{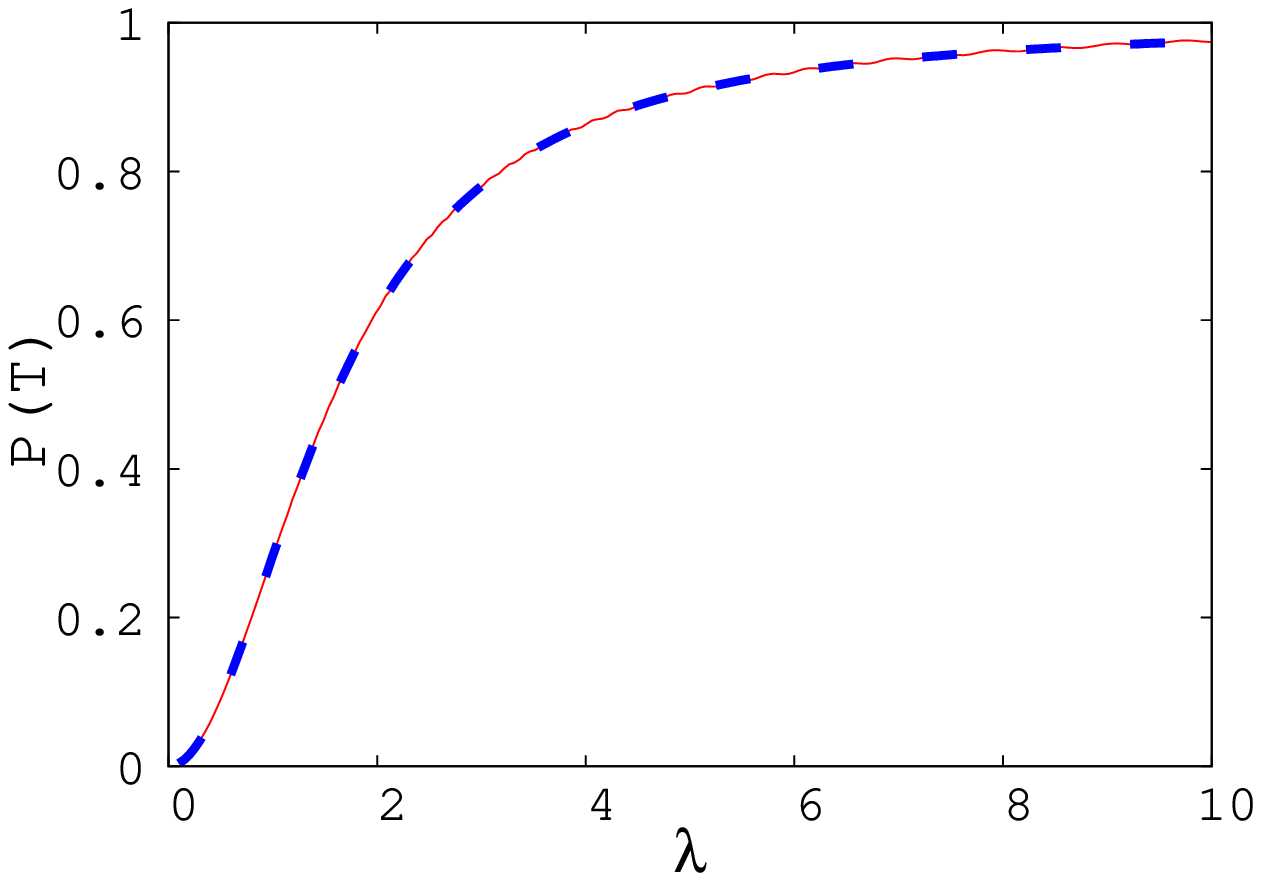}} %
\subfigure[]{\includegraphics[width=0.35\textwidth, angle=0]{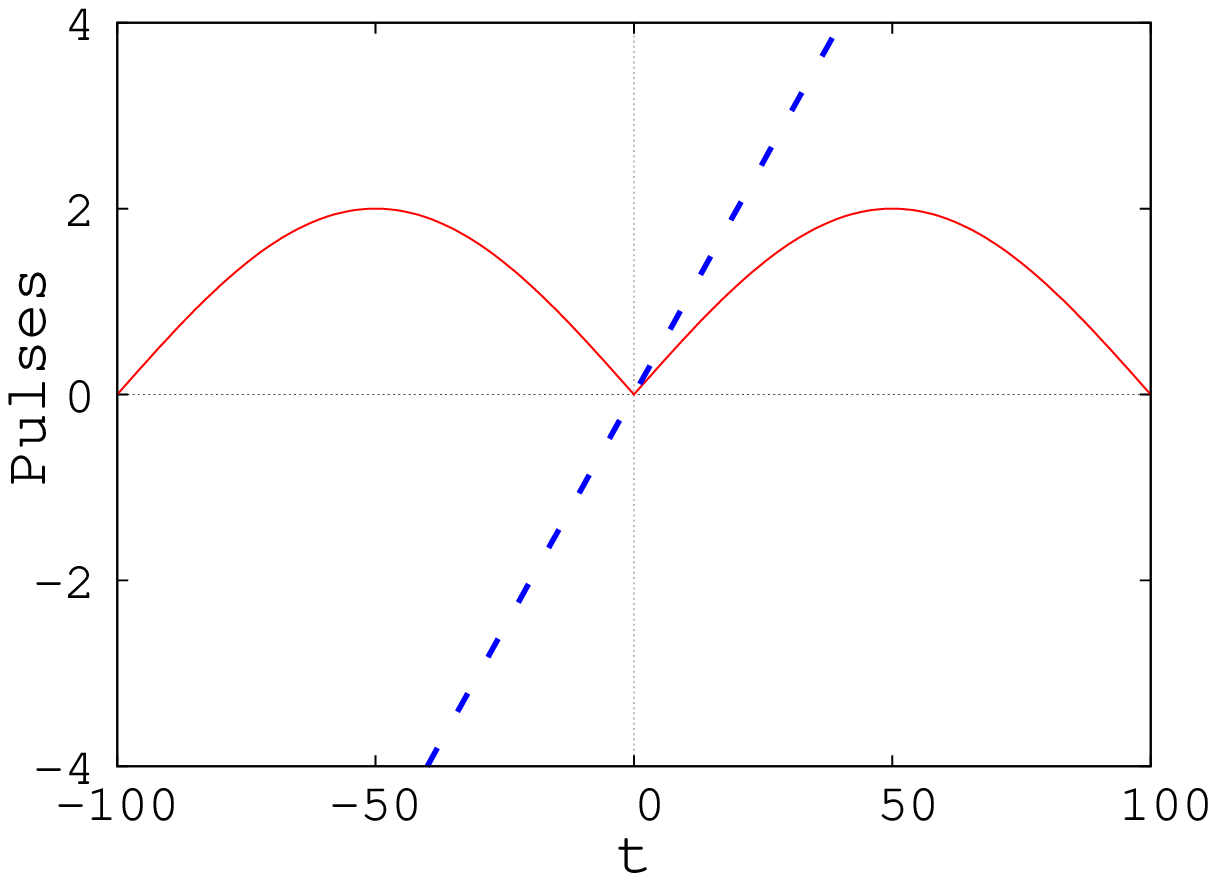}} %
\caption{(Color online) Frame (a) shows the asymptotic populations
of the bare state $\Ket{\uparrow}$ (red solid line) and the
theoretical prediction of the efficiency (blue dashed line),
$\cos^2\Theta$, vs the parameter $\lambda$  (dimensionless). 
Here $\Omega=\lambda\Omega_0\sin(\pi t / T)$, where
$2T$ is the duration of the experiment, and $\Delta=\kappa t$,
with $k/\Omega_0^2 = 0.025$ and $\kappa T/\Omega_0=5$. Frame (b) shows the
coupling $\Omega$ (red solid line) and the detuning $\Delta$ (blue
dashed line) for $\lambda=1$. The amplitudes of the pulses are in
units of $\Omega_0/2$ and time is in units of $2/\Omega_0$.}
\label{fig:Omega_Sin}
\end{figure}

In the standard treatment of \LZ processes, $\Omega$ is assumed to
be constant and $\Delta$ is considered as a linear function of
time. In our examples, both for the sake of simplicity and to make
it easier to compare our situations with the standard one, we have
always assumed $\Delta=\kappa t$, with $\kappa$ being a suitable
parameter.

In Fig.~\ref{fig:Omega_Exp} we show the population transfer
efficiency (i.e., the final population of the initially
unpopulated state) when the $\Omega$-pulse is essentially constant
except for the neighborhood of the level crossing:
$\Omega=\Omega_0(1-\mathrm{exp}(-|t/\sigma|))$. The parameter
$\sigma$ measures the size of the region where $\Omega$ mainly
varies. In Fig.~\ref{fig:Omega_Sin} we consider a case where the
shape of $\Omega$ is far from being constant:
$\Omega=\lambda\Omega_0\sin(\pi t / T)$, where $T$ represents the
duration of the whole process, while $\lambda$ determines the
maximum values of the pulse. In both Figs.~\ref{fig:Omega_Exp} and
\ref{fig:Omega_Sin}, a very good agreement between the results
from the numerical analysis and the theoretical predictions
($\cos^2\Theta$) is evident.

\begin{figure}
\subfigure[]{\includegraphics[width=0.35\textwidth, angle=0]{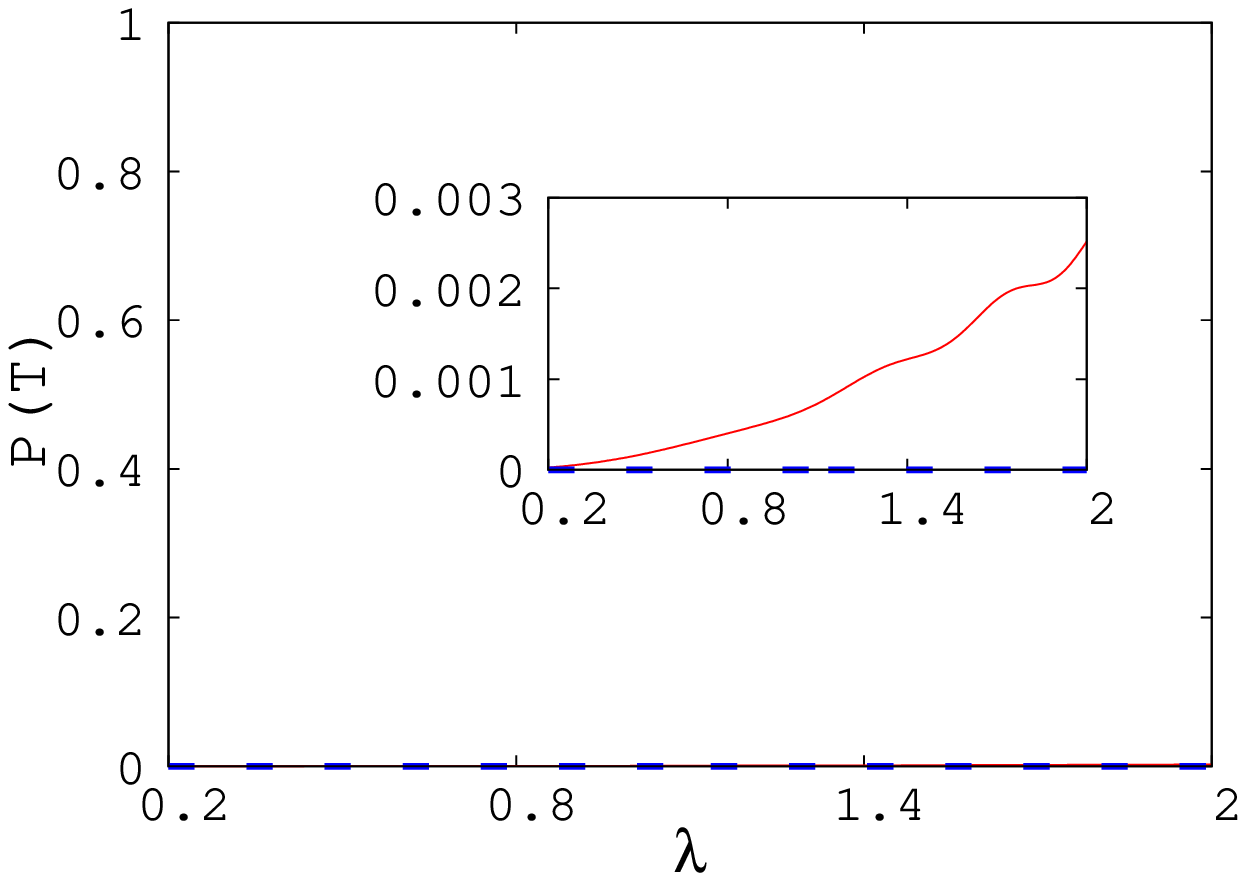}} %
\subfigure[]{\includegraphics[width=0.35\textwidth, angle=0]{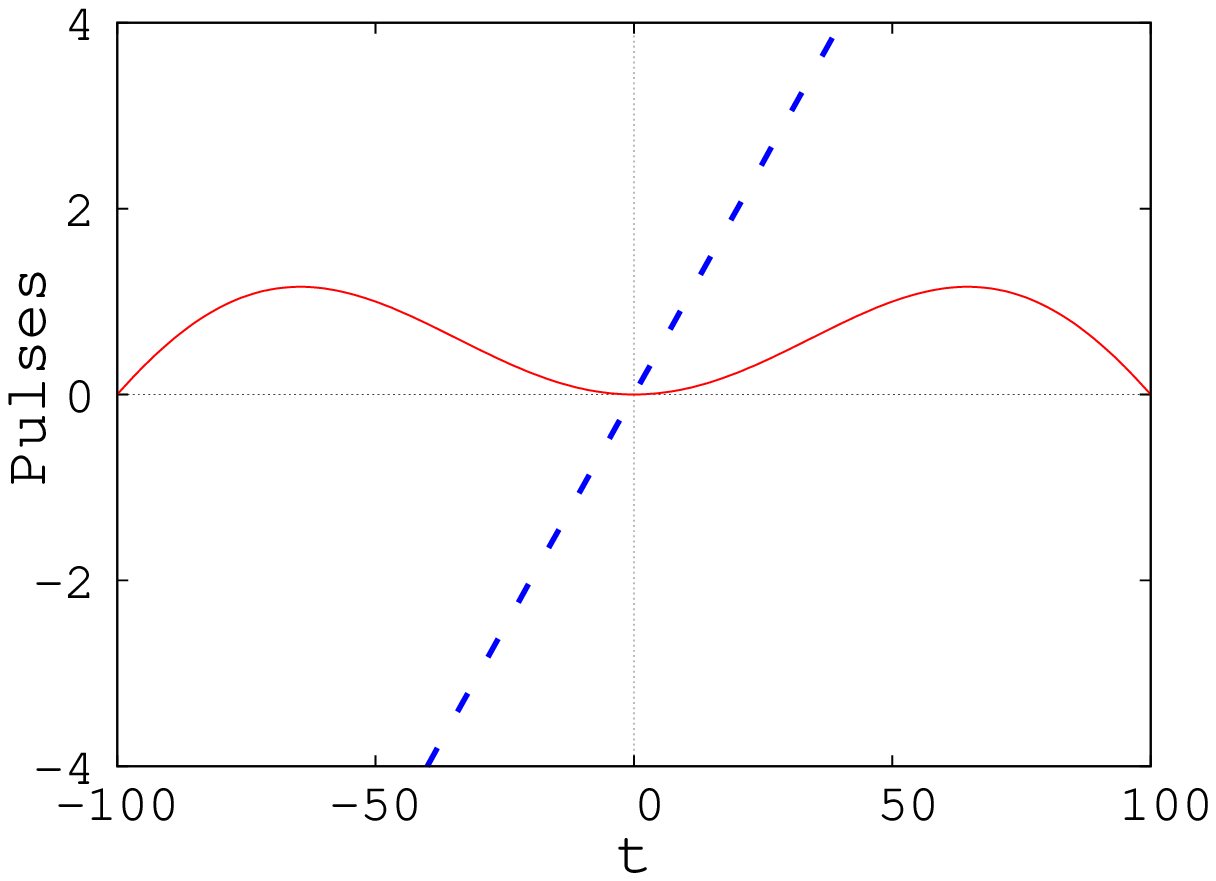}} %
\caption{(Color online) Frame (a) shows the asymptotic populations
of the bare state $\Ket{\uparrow}$ (red solid line) and the
theoretical prediction of the efficiency (blue dashed line),
$\cos^2\Theta$, vs the parameter $\lambda$ (dimensionless). 
The inset is a \lq zoom\rq\, that shows the same
quantities in a different $y$-range. Here $\Omega= \lambda \beta
t \sin(\pi t / T)$, where $2T$ is the duration of the experiment,
and $\Delta=\kappa\,t$, with $\kappa^2/\beta=5$ and $\beta T^2 = 200$. 
Frame (b) shows the coupling $\Omega$ (red solid line) and
the detuning $\Delta$ (blue dashed line) for $\lambda=1$. The
amplitudes of the pulses are in units of $\sqrt{\beta/2}$ and time is
in units of $\sqrt{2/\beta}$.} \label{fig:Omega_TSin}
\end{figure}

Figure~\ref{fig:Omega_TSin} compares numerical and analytical
results for $\Omega= \lambda \beta t \sin(\pi t / T)$. We
observe nearly zero population transfer efficiency, which is due
to the fact that $\alpha = \Delta/\Omega$ diverges at $t=0$,
assuming negative values for $t<0$ and positive values on the
other side, which makes $\theta(0^-)=0$ and $\theta(0^+)=\pi/2$.
The numerical results and theoretical predictions have a very good
agreement also this time.
Of course, in order to support the applicability of our
theoretical analysis, we have always carefully checked (through
numerical evaluation of $\eta$) that the adiabatic approximation
is valid essentially everywhere, with the only possible exception
of $t=0$ and its vicinity.


\subsection{Transitions around the level crossing}

It can happen that the validity of the adiabatic approximation is
broken before reaching the level crossing and for a while after it
has occurred. In such a case, assuming that it is valid out of the
(small) time interval $[-\tau,\tau]$, one can write:
\begin{eqnarray}
\nonumber
U(+\infty,-\infty) &=& U(+\infty,\tau) U(\tau,-\tau)
U(-\tau,-\infty)\\
\nonumber
&\approx&  U_\mathrm{A}(+\infty,\tau) U(\tau,-\tau)
U_\mathrm{A}(-\tau,-\infty)\,. \\
\end{eqnarray}

Since $\tau$ is very small and assuming that $H$ does not exhibit
singularities at any time, we can make the following (rough)
approximation:
\begin{eqnarray}
\nonumber U(\tau,-\tau) &\approx&  \mathbf{I} -\ii
\int_{-\tau}^\tau H(s)\mathrm{d}s\,,\\
&\approx&  \exp\left( -\ii \int_{-\tau}^\tau H(s)\mathrm{d}s
\right)\,.
\end{eqnarray}

If $\Delta$ is an odd function, then one easily obtains:
\begin{equation}
U(\tau,-\tau) \approx \cos\xi\,\mathbf{I} -
\ii\sin\xi\,\sigma_x\,, \qquad \xi =
\int_{-\tau}^\tau\Omega(s)\mathrm{d}s\,.
\end{equation}
Therefore, assuming adiabaticity everywhere out of the interval
$[-\tau,\tau]$, the efficiency of the population transfer could be
roughly given by:
\begin{eqnarray}
\nonumber
&& |\Bra{\psi_\pm(\tau)}U(\tau,-\tau)\Ket{\psi_\pm(-\tau)}|^2 \\
\nonumber && \qquad \approx \cos^2\xi \,
\cos^2[\theta(\tau)-\theta(-\tau)]\\
\label{eq:Eff_with_Psi} && \qquad +\sin^2\xi \,
\sin^2[\theta(\tau)+\theta(-\tau)]\,.
\end{eqnarray}

If the adiabaticity is broken far from the level crossing then all
this treatment is not able to describe the behavior of the system.
On the other hand, even under the proper hypotheses about
adiabaticity, \eqref{eq:Eff_with_Psi} may give a wrong prediction
because the first order approximation to evaluate $U(-\tau,\tau)$
can be insufficient. In this case, further terms of the relevant
Dyson expansion have to be used.

\section{Discussion}\label{sec:discussion}

In this paper we have analyzed the dynamics of a two-state system
whose energies undergo a level crossing. This is a kind of
situation that cannot be treated by the \LZSM~model, because the
interaction between the two states approaches zero when the bare
energies cross.

The theoretical analysis makes it possible to predict the
transition probability from one bare state to the other on the
basis of the way the bare energies and the intensity of the
interaction between them approach zero at the same time. Indeed,
under the assumption that the adiabatic approximation is valid
essentially everywhere except, possibly, around the crossing, the
efficiency is mainly affected by the possible abrupt changes of
the eigenstates of the Hamiltonian. In fact, if the bare energies
cross faster than the interaction there is no abrupt change of the
eigenstates and the population transfer efficiency tends to unity,
while in the opposite case there is the most dramatic abrupt
change of the eigenstates (each eigenstate immediately after the
crossing is orthogonal to the corresponding eigenstate just before
the crossing) and the efficiency tends to zero. If $\Delta$ and
$\Omega$ approach zero as infinitesimal of the same order, then a
specific analysis is required and the exact value of their ratio
determines the efficiency of the population transfer.

Our theoretical analysis is supported by numerical
calculations. We have always found a good agreement between the
theoretically predicted efficiency and the corresponding numerical
result, and in some cases this agreement is perfect.

It is worth noting that in the very neighborhood of the level
crossing nothing really happens, from the dynamical point of view,
in the sense that the state of the quantum system essentially does
not evolve (being $H \approx 0$), but what happens at the level
crossing significantly affects the dynamics from there on. This
circumstance is typical of any sudden change of the Hamiltonian,
but in our case the peculiarity is that such a change is
determined by infinitesimal quantities and the way they compare
each other. Somehow, it resembles a symmetry breaking, since the
perfectly symmetric situation at the crossing ($H=0$) is
significantly altered, with important consequences on the system,
by very small perturbations such as the pulses around $t=0$.

\section*{Acknowledgements}

This work has been financially supported by the Co.R.I. program of
the University of Palermo.


\begin{thebibliography}{99}

\bibitem{ref:Messiah} A. Messiah, \textit{Quantum Mechanics} (Dover)

\bibitem{ref:LZSM-1} L. D. Landau, Phys. Z. Sowjetunion \textbf{2}, 46 (1932).

\bibitem{ref:LZSM-2} C. Zener, Proc. R. Soc. Lond. Ser. A \textbf{137}, 696 (1932).

\bibitem{ref:LZSM-3} E. C. G. Stueckelberg, Helv. Phys. Acta \textbf{5}, 369 (1932).

\bibitem{ref:LZSM-4} E. Majorana, Nuovo Cimento \textbf{9}, 43 (1932).


\bibitem{ref:Nori_review} S.-N. Shevchenko, S. Ashahab, and F. Nori, Phys. Rep. {\bf 492}, 1 (2010).

\bibitem{ref:JJ1} D.M. Berns, {\it et al.}, Nature (London) {\bf 455}, 51 (2008).

\bibitem{ref:JJ2} G. Sun, {\it et al.}, Nature Commun. {\bf 1}, 51 (2010).

\bibitem{ref:BEC} A. Zenesini, D. Ciampini, O. Morsch, and E. Arimondo, Phys. Rev. A {\bf 82}, 065601 (2010).

\bibitem{ref:Nori_spinorial} J.-N. Zhang, C.-P. Sun, S. Yi, and F. Nori, Phys. Rev. A {\bf 83}, 033614 (2011).

\bibitem{ref:Vitanov97b} N. V. Vitanov and B. M. Garraway, Phys. Rev. A \textbf{53}, 4288 (1996); \textit{ibid.} \textbf{54}, 5458(E) (1996).

\bibitem{ref:Vitanov99} N. V. Vitanov, Phys. Rev. A \textbf{59}, 988 (1999).

\bibitem{ref:Vitanov99nonlinear} N. V. Vitanov and K.-A. Suominen, Phys. Rev. A \textbf{59}, 4580 (1999).

\bibitem{ref:Multilevel1} A. V. Shytov, Phys. Rev. A \textbf{70}, 052708 (2004).

\bibitem{ref:Multilevel2} G. S. Vasilev, S. S. Ivanov and N. V. Vitanov, Phys. Rev. A \textbf{75}, 013417 (2007).

\bibitem{ref:Multilevel3} S. S. Ivanov and N. V. Vitanov, Phys. Rev. A \textbf{77}, 023406 (2008).

\bibitem{ref:Nonlinear} A. Ishkhanyan, M. Mackie, A. Carmichael, P. L. Gould, and J. Javanainen, Phys. Rev. A \textbf{69}, 043612 (2004).

\bibitem{ref:Chains1} A. del Campo, M. M. Rams and W. H. Zurek, Phys. Rev. Lett. \textbf{109}, 115703 (2012).

\bibitem{ref:Chains2} N. A. Sinitsyn, Phys. Rev. A \textbf{87}, 032701 (2013).

\bibitem{ref:Zhang2014} J. Zhang, J. Zhang, X. Zhang and K. Kim, Phys. Rev. A \textbf{89}, 013608 (2014).

\bibitem{ref:Gasparinetti2011} S. Gasparinetti, P. Solinas and J. P. Pekola, Phys. Rev. Lett. \textbf{107}, 207002 (2011).

\bibitem{ref:Saquet2010} N. Saquet, A. Cournol, J. Beugnon, J. Robert, P. Pillet and N. Vanhaecke, Phys. Rev. Lett. \textbf{104}, 133003 (2010).

\bibitem{ref:Fishman1990} S. Fishman, K. Mullen and E. Ben-Jacob, Phys. Rev. A \textbf{42}, 5181 (1990).

\bibitem{ref:Bouwmeester1995} D. Bouwmeester, N. H. Dekker, F. E. v. Dorsselear, C. A. Schrama, P. M. Visser and J. P. Woerdman, Phys. Rev. A \textbf{51}, 646 (1995).

\bibitem{ref:shortcuts1} M. V. Berry, J. Phys. A \textbf{42} 365303 (2009).

\bibitem{ref:shortcuts2}  E. Torrontegui, S.Ib\'{a}\~{n}ez, S. Mart\'{i}nez-Garaot, M. Modugno, A. del Campo, D. Gu\'{e}ry-Odelin, A. Ruschhaupt, Xi Chen, J. Gonzalo Muga, Adv. At. Mol. Opt. Phys. \textbf{62}, 117 (2013).

\bibitem{ref:shortcuts3} A. del Campo and M. G. Boshier, Scientific Reports \textbf{2}, 648 (2012).





\end{thebibliography}
\end{document}